\RequirePackage{fix-cm}
\documentclass[smallextended]{svjour3}       
\smartqed  
\usepackage{graphicx}
\usepackage{bm}
\begin{document}

\title{Explanation of Superfluidity Using the Berry Connection for Many-Body Wave Functions
}
\subtitle{ }


\author{Hiroyasu Koizumi 
}


\institute{H. Koizumi \at　Division of Quantum Condensed Matter Physics, 
              Center for Computational Sciences, University of Tsukuba,Tsukuba, Ibaraki 305-8577, Japan \\
              Tel.: +81-29-8536403\\
              Fax: +81-29-8536403\\
              \email{koizumi.hiroyasu.fn@u.tsukuba.ac.jp}           
}

\date{Received: date / Accepted: date}

\maketitle

\begin{abstract}
We show that two phenomena of superfluidity, superfluidity of weakly interacting bosons and superconductivity of the BCS model, are
unified using the collective mode arising from
the Berry connection for many-body wave functions. The superfluidity is attributed to the presence of this mode, which is stabilized by the interaction between particles that causes fluctuations of the number of particles participating in it.
It is suggested that the existence of this collective mode and its stabilization is more fundamental to the occurrence of superconductivity than the electron-pair formation.
\end{abstract}

\section{Introduction}

Properties of superfluid are well described by a macroscopic wave function
\begin{eqnarray}
\Psi_{\rm s}=|\Psi_{\rm s}| e^{ i \varphi}
\label{eq1}
\end{eqnarray}
Thus, the origin of it is very important for the theory of superfluid.

There are two major theories for the definition of $\Psi_{\rm s}$. 
One of them attributes it to the mean value of the boson particle field operator $\hat{\psi}$ \cite{Anderson66},
\begin{eqnarray}
\Psi_{\rm s}=\langle  \hat{\psi} \rangle 
\label{def1}
\end{eqnarray}
In order that this mean value can be nonzero, it has to be calculated by a particle number mixed pure state since $\hat{\psi}$ changes the number of particles by one. It has to be a pure state to explain the observed interference effect attributed to $\Psi_{\rm s}$. 
However, such a state is not realizable if the particle number refers to the total number according to the superselection rule \cite{WWW1970,Peierls1991,Peierls92,LeggettBook}
since the relevant Hamiltonian does not change the particle number. 
Therefore, the particle number in the above is that in the condensate state of a total number fixed system.

The other attributes it to
\begin{eqnarray}
\Psi_{\rm s}=\sqrt{N_0(t)} \eta_0
\label{def2}
\end{eqnarray}
where $N_0$ is the condensate number (the number of particles in the condensate state) and $\eta_0$ is the condensate state single-particle wave function that appears in the natural orbital basis $\{ \eta_i \}$ expression for the one-particle reduced density matrix 
\begin{eqnarray}
\langle  \hat{\psi}^{\dagger}( {\bf r},t) \hat{\psi} ( {\bf r}',t)\rangle =\sum_i N_i(t) \eta_i^{\ast}( {\bf r}) \eta_i( {\bf r}') 
\label{reduceM}
\end{eqnarray}
where $i=0$ state is the condensation state for the Bose-Einstein condensation \cite{Leggett2001,LeggettBook}. This is the one adopted in the Gross-Pitaevskii theory \cite{Gross,Pitaevskii}. This has to be supplemented by the condition for the absence of low energy excitations that will destroy the superfluidity; such a condition is given by Landau \cite{Landau1941}, and the excitation spectrum that satisfies it is provided by Bogoliubov \cite{Bogolubov47}, where the excitation spectrum was calculated using the particle number non-fixed formalism.
Later, the same excitation spectrum was re-derived using the particle number fixed formalism by Leggett \cite{Leggett2001}.

Now let us turn to another superfluid phenomenon, superconductivity of electrons. The macroscopic wave function $\Psi_{\rm GL}$ that appears in the Ginzburg-Landau theory \cite{GL} may be identified as $\Psi_s=\Psi_{\rm GL}$. In superconductivity, $\Psi_s$ is usually attributed to 
\begin{eqnarray}
\Psi_s=\langle \hat{\Psi}_{\uparrow} \hat{\Psi}_{\downarrow} \rangle
\label{psi-s1}
\end{eqnarray}
where $\hat{\Psi}_{\sigma}$ is the field operator for electrons with spin $\sigma$ \cite{Gorkov}. In this definition, the superconducting state is assumed to be a particle number mixed pure state. As in the boson Bose-Einstein case, such a state should be regarded as the state with fluctuating number of particles in the condensate, but with a fixed total number of particles.
 There is an attempt to derive $\Psi_s$ from the total particle number fixed theory by following the method developed for the bosonic superfluity \cite{LeggettBook}. However, the electron pairs do not obey the boson commutation relations, thus, the Bose-Einstein condensation cannot be applied to them. The quantum interference effect indicates that there is a way to compare the phase of $\Psi_s$ at separate coordinate points, and it will be provided by a connection of geometry in mathematics. 

In the present work, we put forward a new approach to define $\Psi_{\rm s}$. It uses the Berry connection for many-body wave function which can be defined for both bosonic and fermionic systems. In this way, $\Psi_s$'s in the above two superfluidity phenomena can be obtained from a single principle. It is also noteworthy that this approach is formulated using fixed total particle number states, thus, is free from artifacts that may arise in the particle number non-fixed formalism. Especially, such a formalism will be useful for the accurate calculations for the properties of nano-sized superconductor systems, including qubits \cite{Nori2017}.

The key ingredient for the particle-number fixed formalism is the number changing operator used in the circuit quantization approach for superconducting qubits \cite{Nori2017}.
In this approach, the Josephson junction is treated as a circuit element with current,
\begin{eqnarray}
I_J=I_c \sin \hat{\phi}
\label{JosephsonC}
\end{eqnarray}
where $I_c$ is a parameter and $\hat{\phi}$ is an operator that corresponds to the scalar $\phi$ satisfying the Josephson relation
\begin{eqnarray}
\dot{\phi}={ {2e} \over \hbar} V(t) 
\label{phi}
\end{eqnarray}
Here, $V(t)$ is the voltage across the junction at time $t$; thus, relations in Eqs.~(\ref{JosephsonC}) and (\ref{phi}) give the I-V relation of the circuit element.

In this approach $\phi$ is a conjugate variable of the Cooper pair number density $\hat{N}_{\rm pair}$, satisfying the commutation relation
\begin{eqnarray}
[\hat{\phi}, \hat{N}_{\rm pair}]= i
\label{comm1}
\end{eqnarray}

Then, $e^{\pm i\hat{\phi}}$ are the number changing operators; i.e., 
if we denote  $|N_{\rm pair} \rangle$ as the eigenstate of $\hat{N}_{\rm pair}$,
\begin{eqnarray}
\hat{N}_{\rm pair}  |N_{\rm pair} \rangle =N_{\rm pair}| N_{\rm pair} \rangle
\end{eqnarray}
 it satisfies
\begin{eqnarray}
e^{\pm i\hat{\phi}} |N_{\rm pair} \rangle =| N_{\rm pair} \mp 1 \rangle
\label{numberOP1}
\end{eqnarray}
This indicates $e^{\pm i\hat{\phi}}$ change the number of the electron-pairs by one.

Strictly speaking, $\hat{\phi}$ is not a hermitian operator \cite{Phase-Angle,Fujikawa2004}; however, we may treat it as hermitian by neglecting a minor difference that affects $N_{\rm pair}=0$ state in the present problem.
 
For the reformulation of superfluid phenomena presented in this work, we use the number changing operators similar to $e^{\pm i\hat{\phi}}$. 
Then, $\Psi_s$ in Eq.~(\ref{psi-s1}) is re-expressed as
\begin{eqnarray}
{\Psi}_s = \langle e^{-i\hat{\phi}} \hat{\Psi}_{\uparrow} \hat{\Psi}_{\downarrow} \rangle
\end{eqnarray}
where the expectation value is evaluated using the particle number fixed state. 
Using the number changing operators defined from the Berry connection, we will provide a novel way of viewing superconductivity.
It will provide the view that the existence of the collective mode from the Berry connection and its stabilization is more fundamental to the occurrence of superconductivity than the electron-pair formation.

The organization of the present work is as follows: in Section~\ref{section2}, the Berry connection for many-body wave functions is defined for boson systems.
In Section~\ref{section3}, the number changing operators are constructed from the Berry connection. 
In Section~\ref{section4}, the Bogoliubov transformations for the interacting boson system is reformulated using the number changing operators. 
In Section~\ref{section5}, the number changing operators for the BCS model are constructed. In Section~\ref{section6}
 the BCS theory is reformulated using the number changing operators.
 In Section~\ref{section7}, the number changing operators obtained in Section~\ref{section5} is identified as the one constructed from the Berry connection for many-body wave functions.
  In Section~\ref{section8} the Josephson tunneling is reformulated using the number changing operators.
   In Section~\ref{section9}, $\Psi_s$ for the BCS model is provided using the number changing operators. 
   In Section~\ref{section10}, the Bogoliubov-de Gennes equations are reformulated using the number changing operators.
Lastly, in Section~\ref{section11}, we conclude the present work with some remarks concerning the foundation of superconductivity.

\section{Berry Connection for Many-Body Wave Functions}
\label{section2}

Let us consider the wave function of a system with spinless $N_b$ bosons,
\begin{eqnarray}
\Phi ({\bf r}_1, \cdots, {\bf r}_{N_b},t)
\label{wavef}
\end{eqnarray}
where ${\bf r}_j$ denotes the coordinate of the $j$th particle.

We define a Berry connection associated with this wave function \cite{Berry,BMKNZ}.
First, we define the parameterized wave function $|n_{\Phi}({\bf r}) \rangle$ with the parameter ${\bf r}$, 
 \begin{eqnarray}
\langle  {\bf r}_{2}, \cdots, {\bf r}_{N_b} |n_{\Phi}({\bf r},t) \rangle = { {\Phi({\bf r}, {\bf r}_{2}, \cdots, {\bf r}_{N_b},t)} \over {|C_{\Phi}({\bf r} ,t)|^{{1 \over 2}}}}
\end{eqnarray}
where $|C_{\Phi}({\bf r} ,t)|$ is the normalization constant given by 
\begin{eqnarray}
|C_{\Phi}({\bf r} ,t)|=\int d{\bf r}_{2} \cdots d{\bf r}_{N_b}\Phi({\bf r}, {\bf r}_{2}, \cdots)\Phi^{\ast}({\bf r},  {\bf r}_{2}, \cdots)
\end{eqnarray}

Using $|n_{\Phi}\rangle$, the {\em Berry connection for many-body wave functions} is defined as
 \begin{eqnarray}
{\bf A}^{\rm MB}_{\Phi}({\bf r},t)=-i \langle n_{\Phi}({\bf r},t) |\nabla_{\bf r}  |n_{\Phi}({\bf r},t) \rangle
\end{eqnarray}
Here, ${\bf r}$ is regarded as the parameter \cite{Berry}. 

We only consider the case where the origin of ${\bf A}_{\Phi}^{\rm MB}$ is not the ordinary magnetic field one; thus, we have 
\begin{eqnarray}
\nabla \times {\bf A}^{\rm MB}_{\Phi}=0
\label{BMB}
\end{eqnarray}
Then, it can be written in the pure gauge form,
\begin{eqnarray}
 {\bf A}^{\rm MB}_{\Phi}=\nabla \varphi
 \label{varphi1}
\end{eqnarray}
where $\varphi$ is a function which may be multi-valued.

The kinetic energy part of the Hamiltonian is given by
\begin{eqnarray}
K_0={ 1\over {2m}} \sum_{j=1}^{N_b} \left( {\hbar \over i} \nabla_{j} \right)^2
\label{a2}
\end{eqnarray}
where $m$ is the particle mass and $\nabla_{j}$ is the gradient operator with respect to the $j$th electron coordinate ${\bf r}_j$.

Using $\Phi$ and ${\bf A}_{\Phi}^{\rm MB}$, we can construct a currentless wave function $\Phi_0$ for the current operator associated with $K_0$
\begin{eqnarray}
\Phi_0 ({\bf r}_1, \cdots, {\bf r}_{N_b},t)=\Phi ({\bf r}_1, \cdots, {\bf r}_{N_b},t)\exp\left(- i \sum_{j=1}^{N_b} \int_{0}^{{\bf r}_j} {\bf A}_{\Phi}^{\rm MB}({\bf r}',t) \cdot d{\bf r}' \right)
\label{wavef0}
\end{eqnarray}

Reversely, $\Phi ({\bf r}_1, \cdots, {\bf r}_{N_b},t)$ is expressed as
 \begin{eqnarray}
\Phi =\Phi_0\exp\left( i \sum_{j=1}^{N_b} \varphi ({\bf r}_j, t) \right)
\label{f}
\end{eqnarray}
using the currentless wave function $\Phi_0$.

\section{The Number Changing Operator}
\label{section3}

Let us obtain the conjugate momentum of $\varphi$ given in Eqs.~(\ref{varphi1}) and (\ref{f}). 
 For this purpose, we use the time-dependent variational principle using the following Lagrangian \cite{Koonin1976},
\begin{eqnarray}
{\cal L}\!=\langle \Phi | i\hbar \partial_t \!-\!H| \Phi \rangle\!=\! i\hbar \langle \Phi_0 | \partial_t | \Phi_0 \rangle- \hbar \int \!d{\bf r} \ {{\rho_b \dot{\varphi} }} - \langle \Phi |H| \Phi \rangle
\label{L}
\end{eqnarray}
where $\rho_b$ is the number density of the bosons.

From the above Lagrangian, the conjugate momentum of $\varphi$ is obtained as
\begin{eqnarray}
p_{\varphi}= {{\delta {\cal L}} \over {\delta \dot{\varphi}}}=-\hbar \rho_b
\label{momentumchi}
\end{eqnarray}
thus, $\varphi$ and $\rho_b$ are canonical conjugate variables.

If we follow the canonical quantization condition $[\hat{p}_{\varphi}({\bf r}, t), \hat{\varphi}({\bf r}', t)]=-i\hbar \delta ({\bf r}- {\bf r}')$, where $\hat{p}_{\varphi}$ and $\hat{\varphi}$ are operators corresponding to ${p}_{\varphi}$ and ${\varphi}$ respectively, 
we have
\begin{eqnarray}
\left[{ {\hat{\rho}_b({\bf r}, t)} } , \hat{\varphi}({\bf r}', t) \right]=i \delta ({\bf r}- {\bf r}')
\label{commu0}
\end{eqnarray}
where $\hat{\rho}_b$ is the operator corresponding to $\rho_b$.
 Strictly speaking, $\hat{\varphi}$ is not a hermitian operator; however, it is known that when it is used as 
 $\sin \hat{\varphi}$ or  $\cos \hat{\varphi}$, the problem is avoided, practically \cite{Phase-Angle}. In the following we use
 $\hat{\varphi}$  as $e^{\pm i {\hat{\varphi} }}$.
 
We construct the following boson field operators from $\hat{\varphi}$ and $\hat{\rho}_b$,
\begin{eqnarray}
\hat{\psi}_b^{\dagger}({\bf r})= \left(\hat{\rho}_b({\bf r}) \right)^{1/2} e^{- i {\hat{\varphi}({\bf r}) }}, \quad \hat{\psi}_b({\bf r})= e^{i { \hat{\varphi}}}\left( \hat{\rho}_b({\bf r}) \right)^{1/2}
\label{boson1}
\end{eqnarray}

Using Eq.~(\ref{commu0}), the following relations are obtained,
\begin{eqnarray}
 [\hat{\psi}_b({\bf r}),\hat{\psi}_b^{\dagger}({\bf r}')]=\delta({\bf r}-{\bf r}'), \quad  [{\psi}_b({\bf r}),\hat{\psi}_b({\bf r}')]=0, \quad  [\hat{\psi}_b^{\dagger}({\bf r}),\hat{\psi}_b^{\dagger}({\bf r}')]=0
 \label{commu2}
\end{eqnarray}

Now, we construct the number changing operators for particles participating in the collective mode described by $\varphi$.
For that purpose, we first define the following creation and annihilation operators,
\begin{eqnarray}
 \hat{B}^{\dagger}=\int_{\cal V} d{\bf r} \hat{\psi}_b^{\dagger}({\bf r}) 
, \quad \hat{B}=\int_{\cal V } d{\bf r} \hat{\psi}_b^{}({\bf r}),  
\label{B}
 \end{eqnarray}
 where ${\cal V}$ is the volume of the system. 
  
Using Eq.~(\ref{commu2}), the following boson commutation relations are obtained,
 \begin{eqnarray}
 \quad [\hat{B}, \hat{B}^{\dagger}]=1, \quad  \quad [\hat{B}, \hat{B}]=0, \quad [\hat{B}^{\dagger}, \hat{B}^{\dagger}]=0
 \label{1}
 \end{eqnarray}
 
 Then, the number operator for the particle participating in the collective mode described by $\varphi$ is given by
 \begin{eqnarray}
\hat{N}_{\varphi}= \hat{B}^{\dagger}  \hat{B}
 \end{eqnarray}
 
We define eigenstates of $\hat{N}_{\varphi}$,
    \begin{eqnarray}
\hat{N}_{\varphi} | N_{\varphi}\rangle =N_{\varphi}| N_{\varphi} \rangle
 \end{eqnarray}
 
 Through the creation and annihilation operators $\hat{B}^{\dagger}$ and $\hat{B}$, the phase operator $\hat{\Theta}$ that is conjugate to the number operator $\hat{N}_{\varphi}$ can be defined,
  \begin{eqnarray}
 \hat{B}^{\dagger}=(\hat{N}_{\varphi})^{ 1 \over 2} e^{ -{ i } \hat{\Theta}}
, \quad \hat{B}=e^{{ i } \hat{\Theta}}(\hat{N}_{\varphi})^{ 1 \over 2} 
\label{chij1}
 \end{eqnarray}
 
The following relation is obtained from Eqs.~(\ref{1}) and (\ref{chij1}),
   \begin{eqnarray}
[e^{ i  \hat{\Theta}}, \hat{N}_{\varphi}]=e^{ { i } \hat{\Theta}}
\label{commchi}
 \end{eqnarray}
 Then, $e^{ { \pm i } \hat{\Theta}}$ are the number changing operators that satisfy
    \begin{eqnarray}
  e^{\pm i \hat{\Theta}} | N_{\varphi} \rangle = | N_{\varphi} \mp 1 \rangle
\label{commchi2}
 \end{eqnarray}

 With the number changing operators, the definition corresponding to Eq.~(\ref{def1}) is derived in a particle number-fixed manner
 from  one-particle reduced density matrix in Eq.~(\ref{reduceM}),
 \begin{eqnarray}
\langle  \hat{\psi}_b^{\dagger} ({\bf r}) \hat{\psi}_b ({\bf r}')\rangle =\langle  \hat{\psi}_b^{\dagger} ({\bf r})   e^{ i \hat{\Theta}}   e^{- i \hat{\Theta}} \hat{\psi}_b ({\bf r}')\rangle \approx \langle  \hat{\psi}_b^{\dagger} ({\bf r})   e^{ i \hat{\Theta}}  \rangle \langle e^{- i \hat{\Theta}} \hat{\psi}_b ({\bf r}')\rangle
\end{eqnarray}
 where $ \hat{\psi}_b^{\dagger} ({\bf r})   e^{ i \hat{\Theta}} $ and $e^{- i \hat{\Theta}} \hat{\psi}_b ({\bf r}')$ conserve the particle number. 

Then, we can define 
\begin{eqnarray}
\Psi^b_s=\langle e^{- i \hat{\Theta}} \hat{\psi}_b ({\bf r})\rangle
\label{Psis}
\end{eqnarray}
for the macroscopic wave function corresponding to $\Psi_s$.

The definition corresponding to Eq.~(\ref{def2}) is derived from the above definition in the following way:
 first we obtain $\varphi({\bf r})$ from Eq.~({\ref{varphi1}),
 \begin{eqnarray}
 \varphi({\bf r})=\int^{\bf r}  {\bf A}^{\rm MB} ({\bf r}') \cdot d{\bf r}'
 \end{eqnarray}
 Next, we define $\eta_0$ as 
  \begin{eqnarray}
 \eta_0({\bf r})=\sqrt{ { \rho_b({\bf r})}  \over {N_{\varphi} }} e^{i \varphi({\bf r})}
 \end{eqnarray}
 
 We construct the basis $\{ \eta_i \}$ where $\eta_0$ is given above, and $\eta_i, i \neq 0$ are chosen to be orthogonal to $\eta_0$. 
Then, $\hat{\psi}_b$ is given by
  \begin{eqnarray}
 \hat{\psi}_b({\bf r})=\sum_i \eta_i ({\bf r}) b_i 
 \end{eqnarray}
 where $b_i$ is the boson annihilation operator for the state $\eta_i$.  
 
A prerequisite to superfluidity for the boson system is the Bose-Einstein condensation. We identify $\eta_0$ as the condensation state. 
Then, we may employ the following approximations,
\begin{eqnarray}
 e^{- i \hat{\Theta}} \approx (\hat{N}_{\varphi})^{-1/2} b_0^{\dagger}, \quad
 \langle b_0^{\dagger} b_0 \rangle =N_0 \approx N_{\varphi}
 \end{eqnarray}
 
 Consequently, we have
\begin{eqnarray}
\Psi^b_s &=& \langle e^{- i \hat{\Theta}} \hat{\psi}_b ({\bf r})\rangle \approx \langle (\hat{N}_{\varphi})^{-1/2} b_0^{\dagger} \sum_i \eta_i ({\bf r}) b_i \rangle
\nonumber
\\
&=&\langle ({N}_{\varphi})^{-1/2}  \eta_0 ({\bf r}) b_0^{\dagger} b_0 \rangle  \approx ({N}_0)^{1/2}\eta_0 ({\bf r}) 
\end{eqnarray}
in accordance with Eq.~(\ref{def2}).

It is worth noting that the present formalism also has a connection to the quantization of liquid motion given by Landau \cite{Landau1941}. He derived the following commutation relation
\begin{eqnarray}
[\hat{\bf v}({\bf r}), m\hat{\rho}_b({\bf r}')]=-i \hbar \nabla\delta ({\bf r} - {\bf r}')
\end{eqnarray}
where $m$ and $\hat{\bf v}$ are the mass and velocity operator, respectively. This relation can be obtained from Eq.~(\ref{commu0}) by identifying $\hat{\bf v}$ to
\begin{eqnarray}
\hat{\bf v}({\bf r})={\hbar \over m}  \nabla \hat{\varphi}({\bf r})
\end{eqnarray}

As shown above, three different previous approaches for superfluidity can be unified by utilizing the Berry connection ${\bf A}_{\Phi}^{\rm MB}$. 

\section{Bogoliubov Operators Using the Number Changing Operators $e^{\pm i \hat{\Theta}}$ }
\label{section4}
 
 The state with the wave function $\Phi$ in Eq.~(\ref{wavef}) may not be the ground state. We consider an interaction between the particles and show that a lower energy state is obtained
 by taking into account fluctuations of the number of particles participating in the collective mode described by $\varphi$.
 For this purpose, we use the model considered by Bogoliubov \cite{Bogolubov47}; 
 we follow the derivation by Bogoliubov and reformulate it using the number changing operator and particle number conserving version of Bogoliubov operators.
  
  The Hamiltonian considered by Bogoliubov is given by
 \begin{eqnarray}
H= \sum_{\bf k} \varepsilon_{\bf k} b_{\bf k}^{\dagger}b_{\bf k}+ { {g} \over {2 {\cal V}}} \sum_{{\bf k}_1, {\bf k}_2, {\bf k}_3,{\bf k}_4} b_{{\bf k}_1}^{\dagger}b_{{\bf k}_2}^{\dagger}b_{{\bf k}_3}b_{{\bf k}_4}\delta_{{\bf k}_1+{\bf k}_2= {\bf k}_3+{\bf k}_4}
\end{eqnarray}
where ${\bf k}$ is the wave vector, $\varepsilon_{\bf k}={{\hbar^2 k^2} \over {2m}}$ is the single-particle energy for the particle with wave vector ${\bf k}$, $g >0$ is the coupling constant for the two-body interaction, and  $b_{\bf k}$ and $b^{\dagger}_{\bf k}$ are boson annihilation and creation operators for the particle with wave vector ${\bf k}$, respectively.

Assuming that the contribution from the second term is much smaller than that from the first one, the zeroth order state is
given as the state with the total occupation of ${\bf k}=0$ state. We take this state as the Bose-Einstein condensation state.

Then, the number changing operator is given by
\begin{eqnarray}
e^{- i \hat{\Theta}}= ({b_0^{\dagger}b_0})^{-1/2} b^{\dagger}_0
\end{eqnarray}

In the interaction term, important terms are those with four ${\bf k}=0$ and two ${\bf k}=0$ terms;
thus, we may approximate it as
\begin{eqnarray}
 { {g} \over {2 {\cal V}}}  \left\{b_{0}^{\dagger}b_{0}^{\dagger}b_{0}b_{0}
+\sum_{{\bf k}\neq0} [2(b_{{\bf k}}^{\dagger}b_{{\bf k}}b_{0}^{\dagger}b_{0}
+b_{-{\bf k}}^{\dagger}b_{-{\bf k}}b_{0}^{\dagger}b_{0}) + b_{{\bf k}}^{\dagger}b_{-{\bf k}}^{\dagger}b_{0}b_{0}
+b_{0}^{\dagger}b^{\dagger}_{0}b_{{\bf k}}b_{-{\bf k}})
]
\right\}
\nonumber
\\
.
\end{eqnarray}

We can also use the following approximations,
\begin{eqnarray}
b_{0}^{\dagger}b_{0}^{\dagger}b_{0}b_{0} \rightarrow N_0(N_0-1)\approx N_0^2, \quad b_{0}^{\dagger}b_{0}  \rightarrow N_0
\end{eqnarray}

We also use the the following replacements
\begin{eqnarray}
b_{{\bf k}}^{\dagger}b_{-{\bf k}}^{\dagger}b_{0}b_{0}&=&b_{{\bf k}}^{\dagger}e^{ i \hat{\Theta}}e^{- i \hat{\Theta}}b_{0}b_{-{\bf k}}^{\dagger}e^{ i \hat{\Theta}}e^{- i \hat{\Theta}}b_{0} \rightarrow N_0 b_{{\bf k}}^{\dagger}e^{ i \hat{\Theta}}b_{-{\bf k}}^{\dagger}e^{ i \hat{\Theta}}
\nonumber
\\
b_{0}^{\dagger}b^{\dagger}_{0}b_{{\bf k}}b_{-{\bf k}}&=&b_{0}^{\dagger}e^{ i \hat{\Theta}}e^{ -i \hat{\Theta}}b_{{\bf k}}b^{\dagger}_{0}e^{ i \hat{\Theta}}e^{- i \hat{\Theta}}b_{-{\bf k}} \rightarrow N_0 e^{- i \hat{\Theta}}b_{{\bf k}} e^{- i \hat{\Theta}}b_{-{\bf k}} 
\end{eqnarray}
where $e^{- i \hat{\Theta}}b_{0}$ and $b_{0}^{\dagger}e^{ i \hat{\Theta}}$ are replaced by their average values $\sqrt{N_0}$.

As a consequence, the Hamiltonian becomes
\begin{eqnarray}
H &\approx&  \sum_{\bf k} \varepsilon_{\bf k} b_{\bf k}^{\dagger}e^{ i \hat{\Theta}}e^{- i \hat{\Theta}}b_{\bf k} \!+\! { {g} \over {2 {\cal V}}}  \left\{ N_0^2
\!+\!N_0\sum_{{\bf k}\neq0} [2(b_{{\bf k}}^{\dagger}e^{ i \hat{\Theta}}e^{- i \hat{\Theta}}b_{{\bf k}}
\!+\!b_{-{\bf k}}^{\dagger}e^{ i \hat{\Theta}}e^{- i \hat{\Theta}}b_{-{\bf k}}) \!+\! b_{{\bf k}}^{\dagger}e^{ i \hat{\Theta}}b_{-{\bf k}}^{\dagger}e^{ i \hat{\Theta}}
\!+\!e^{- i \hat{\Theta}}b_{{\bf k}}e^{- i \hat{\Theta}}b_{-{\bf k}})
]
\right\}
\nonumber
\\
 &\approx& { {g N_0^2} \over {2 {\cal V}}} \!+\! 
 \sum_{{\bf k}\neq 0}  \left\{ \left( { \varepsilon_{\bf k} \over 2} \!+\! { {g N_0} \over { {\cal V}}} \right) (b_{{\bf k}}^{\dagger}e^{ i \hat{\Theta}}e^{- i \hat{\Theta}}b_{{\bf k}}
\!+\!b_{-{\bf k}}^{\dagger}e^{ i \hat{\Theta}}e^{- i \hat{\Theta}}b_{-{\bf k}}) \!+\!  \left( { {g N_0} \over { 2{\cal V}}} \right)(b_{{\bf k}}^{\dagger}e^{ i \hat{\Theta}}b_{-{\bf k}}^{\dagger}e^{ i \hat{\Theta}}
\!+\!e^{- i \hat{\Theta}}b_{{\bf k}}e^{- i \hat{\Theta}}b_{-{\bf k}})
\right\}
\end{eqnarray}

We further replace $N_0$ by using the total number operator 
\begin{eqnarray}
\hat{N}=N_0+{1 \over 2}  \sum_{{\bf k}\neq 0}(b_{{\bf k}}^{\dagger}e^{ i \hat{\Theta}}e^{- i \hat{\Theta}}b_{{\bf k}}
+b_{-{\bf k}}^{\dagger}e^{ i \hat{\Theta}}e^{- i \hat{\Theta}}b_{-{\bf k}})
\end{eqnarray}
and approximate $\hat{N}$ by $N$.

Then, by keeping terms of order $N^2$ and $N$, we have
\begin{eqnarray}
H \approx { {g N^2} \over {2 {\cal V}}} \!+\! { 1\over 2}
 \sum_{{\bf k}\neq 0}  \left\{ \left( { \varepsilon_{\bf k}} \!+\! { {g N} \over { {\cal V}}} \right) (b_{{\bf k}}^{\dagger}e^{ i \hat{\Theta}}e^{- i \hat{\Theta}}b_{{\bf k}}
\!+\!b_{-{\bf k}}^{\dagger}e^{ i \hat{\Theta}}e^{- i \hat{\Theta}}b_{-{\bf k}}) \!+\!  \left( { {g N} \over { {\cal V}}} \right)(b_{{\bf k}}^{\dagger}e^{ i \hat{\Theta}}b_{-{\bf k}}^{\dagger}e^{ i \hat{\Theta}}
\!+\!e^{- i \hat{\Theta}}b_{{\bf k}}e^{- i \hat{\Theta}}b_{-{\bf k}})
\right\}
\nonumber
\\
\end{eqnarray}

Now we use the following particle number conserving version of Bogoliubov operators
\begin{eqnarray}
\beta_{\bf k}&=&u^b_k e^{ -i \hat{\Theta}}b_{\bf k} + v^b_k b^{\dagger}_{-{\bf k}} e^{ i \hat{\Theta}}
\nonumber
\\
\beta^{\dagger}_{-{\bf k}}&=&v^b_k b^{\dagger}_{-{\bf k}} e^{ i \hat{\Theta}}+ u^b_k e^{ -i \hat{\Theta}}b_{\bf k}
\label{BogoliubovO}
\end{eqnarray}
They satisfy the boson commutation relations; parameters $v^b_k$ and $u^b_k$ are given by
\begin{eqnarray}
(v^b_k)^2 = (u^b_k)^2-1={1 \over 2} \left[(E^b)^{-1}_k \left( { \varepsilon_{\bf k}} + { {g N} \over {\cal V}} \right)-1 \right]
\end{eqnarray}
where
$E^b_k$ is 
\begin{eqnarray}
E^b_k = \sqrt{  \left( \varepsilon_{\bf k}+ { {g N} \over { {\cal V}}} )^2-({ {g N} \over  {\cal V}} \right)^2}
\end{eqnarray}

Then, the Hamiltonian is expressed as
\begin{eqnarray}
H \approx
{ 1\over 2}
 \sum_{{\bf k}\neq 0} E^b_k (\beta_{{\bf k}}^{\dagger}\beta_{{\bf k}}
+\beta_{-{\bf k}}^{\dagger} \beta_{-{\bf k}}) +
 { {g N^2} \over {2 {\cal V}}} - { 1\over 2}
 \sum_{{\bf k}\neq 0} \left( { \varepsilon_{\bf k}} + { {g N} \over { {\cal V}}} -E^b_k\right) 
 \label{BoseH}
 \end{eqnarray}

The ground state $|{\rm Gnd}^b \rangle$ is defined by the condition
\begin{eqnarray}
\beta_{\bf k} |{\rm Gnd}^b \rangle =0 \mbox{  for ${\bf k}\neq0$ }
\label{ground}
\end{eqnarray}

The equation (\ref{BoseH}) indicates that the ground state is energetically lowered by the interaction that causes fluctuations of the number of particles participating in the condensate.
This situation is similar to the one find in the antiferromagnetic ground state, where the semiclassical equation-of-motion method \cite{Anderson1952} and 
the Holstein-Primakoff method \cite{Kubo1952} that take into account spin-waves yield the lower energy state than the N\'{e}el state. 
In the present case, the fluctuations of the number of particles participating in the condensate plays a role of the spin-waves.
It is also noteworthy that a similar view is also provided for the BCS superconducting state using the random-phase approximation \cite{Anderson1958b}.

 The ground state that satisfies Eq.~(\ref{ground}) is given by
\begin{eqnarray}
|{\rm Gnd}^b (N)\rangle = \prod_{{\bf k} \neq 0} \left( u^b_{k} -v^b_{k}b^{\dagger}_{\bf k}b^{\dagger}_{-{\bf k}} e^{2i\hat{\Theta}} \right)|{\rm Cnd}^b(N) \rangle 
\label{Gndb}
\end{eqnarray}
where the condensate state is given by
\begin{eqnarray}
|{\rm Cnd}^b(N)\rangle = e^{-iN\hat{\Theta}}|{\rm vac} \rangle 
\label{Cnd0}
\end{eqnarray}
and $|{\rm vac} \rangle $ denotes the vacuum. The number of particles in the condensate fluctuates due to the number changing operator $e^{2i\hat{\Theta}}$ in Eq.~(\ref{Gndb}). 

The above ground state is actually equivalent to the one obtained by Leggett \cite{LeggettBook,Leggett2001},
\begin{eqnarray}
|{\rm Gnd}^b(N) \rangle ={\rm const.} \left( b_0^{\dagger} b^{\dagger}_0 - \sum_{{\bf k}\neq 0} { v^b_k \over u^b_k} b^{\dagger}_{\bf k}
b^{\dagger}_{-{\bf k}}  \right)^{N/2}|{\rm vac} \rangle 
\end{eqnarray}

Now the excited state is given by
\begin{eqnarray}
\beta^{\dagger}_{\bf k}|{\rm Gnd}^b(N) \rangle 
\end{eqnarray}
with excitation energy $E^b_k$. The single-particle dispersion $E^b_k$ satisfies the condition provided by Landau \cite{Landau1941} for the superfluidity around the zero excitation energy. 

From the point of view of the present theory, the definition of $\Psi_s$ in Eq.~(\ref{def1}) (which corresponds to Eq.~(\ref{Psis}) in the present theory) may be regarded that superfluidity is attributed to the stabilization of the collective mode. This stabilization requires the fluctuation of the number of particles participating in the collective mode.

On the other hand, the definition in Eq.~(\ref{def2}) attributes the superfluidity to the existence of the condensate state with the collective mode ($|{\rm Cnd}^b(N)\rangle$ in Eq.~(\ref{Cnd0})); without it, the superfluidity dose not arise from the beginning.

The present formalism combines the above two views; namely, the superfluidity requires both the existence of the condensate state with the collective mode, and the stabilization of the collective mode by the fluctuation of the number of particles participate in the collective mode.

\section{Number Changing Operators for the BCS Model}
\label{section5}

In this section, we obtain the number changing operators for superconductivity.
First, we review the BCS theory \cite{BCS1957}, succinctly. The model Hamiltonian is given by $H_{\rm BCS}=H_{\rm kin}+H_{\rm int}$; 
$H_{\rm kin}$ is the kinetic energy 
\begin{eqnarray}
H_{\rm kin}= \sum_{{\bf k} \sigma} \xi_0({\bf k})c^{\dagger}_{{\bf k} \sigma}c_{{\bf k} \sigma}
\end{eqnarray}
where $c^{\dagger}_{{\bf k} \sigma}$ and $c_{{\bf k} \sigma}$ are annihilation and creation operators for electrons with wave vector ${\bf k}$ and spin $\sigma$, respectively, and $\xi_0({\bf k})$ is the single-particle energy measured from the Fermi energy ${\cal E}_{F}$,
\begin{eqnarray}
\xi_0({\bf k})={\cal E}({\bf k})-{\cal E}_{F}
\end{eqnarray}
$H_{\rm int}$ is the interaction energy given by
\begin{eqnarray}
H_{\rm int}=\sum_{{\bf k} {\bm \ell}} V_{{\bf k} {\bm \ell}}c^{\dagger}_{{\bf k} \uparrow}c^{\dagger}_{-{\bf k} \downarrow}c_{-{\bm \ell} \downarrow}c_{{\bm \ell} \uparrow}.
\label{Hint}
\end{eqnarray}
We assume the simplest form for $V_{{\bf k} {\bm \ell}}$, $V_{{\bf k} {\bm \ell}}=-g <0$, where $g$ is a constant.  

 The superconducting state is given by the following state vector,
\begin{eqnarray}
|{\rm BCS} \rangle=\prod_{\bf k}(u_{k}+v_{k}c^{\dagger}_{{\bf k} \uparrow}c^{\dagger}_{-{\bf k} \downarrow})
|{\rm vac} \rangle.
\label{BCS}
\end{eqnarray}
This state exploits the attractive interaction between electron pairs $({\bf k} \uparrow)$ and $(-{\bf k} \downarrow)$, and gives rise to the energy gap 
\begin{eqnarray}
\Delta_{\rm BCS}=g\sum_{{\bm \ell} } u_{\bm \ell}v_{\bm \ell}
\end{eqnarray}
where
 $u_{\bf k}$ and $v_{\bf k}$ are parameters given by
\begin{eqnarray}
u_{k}&=&{1 \over \sqrt{2}} \left(1 + {{\xi_0({\bf k})} \over E_{\bf k} } \right)^{1/2}
\nonumber
\\
v_{k}&=&{1 \over \sqrt{2}} \left(1 -{{\xi_0({\bf k})} \over E_{\bf k} } \right)^{1/2},
\label{uv}
\end{eqnarray}
and $E_{\bf k}$ is the Bogoliubov excitation energy
\begin{eqnarray}
E_{\bf k} = [ \Delta_{\rm BCS}^2 + \xi^2_0({\bf k})]^{1/2}
\end{eqnarray}

For the BCS model, a relation similar to Eq.~(\ref{commu0}) is obtained as follows: first, we divide the system into coarse-grained cells of unit volumes; and express the BCS state in the coarse-grained cell with the center position ${\bf r}$ as
\begin{eqnarray}
|\Psi_{\rm BCS} ({\bf r},t) \rangle=\prod_{\bf k} \left( u_{k} ({\bf r},t) + e^{-i \chi({\bf r},t) } v_{k} ({\bf r},t) c^{\dagger}_{{\bf k} \uparrow}c^{\dagger}_{-{\bf k} \downarrow} \right) |{\rm vac} \rangle
\label{BCSr}
\end{eqnarray}
Now $u_k$ and $v_k$ depend on the coordinate and time. The phase factor $e^{-i \chi({\bf r},t) }$ is the phase of the order parameter.

Then, the Lagrangian corresponding to Eq.~(\ref{L}) is given by
\begin{eqnarray}
{\cal L}_{\rm BCS}\!=\int d{\bf r} \langle \Psi_{\rm BCS} ({\bf r},t)  | i\hbar \partial_t \!-\!H_{\rm BCS}| \Psi_{\rm BCS} ({\bf r},t)  \rangle\!=\!\int \!d{\bf r} \ {{\rho_{e} \dot{\chi} \hbar} \over 2}\!-\! \int d{\bf r}\langle \Psi_{\rm BCS}({\bf r},t)  | H_{\rm BCS} | \Psi_{\rm BCS} ({\bf r},t)  \rangle\!
\nonumber
\\
\label{L2}
\end{eqnarray}
where $H_{\rm BCS}$ is now coordinate dependent and specified by the coarse-grained cell position ${\bf r}$.
The electron number density in the cell is given by
\begin{eqnarray}
\rho_{e}({\bf r},t)= 2\sum_{\bf k} u_{k} ({\bf r},t) v_{k} ({\bf r},t)
\label{rhoBCS}
\end{eqnarray}

Using the Lagrangian in Eq.~(\ref{L2}), we obtain $p_{\chi}=\hbar \rho_{e} /2$; thus, we have
\begin{eqnarray}
\left[ \hat{\rho}_{e}({\bf r}, t), {\hat{\chi}({\bf r}', t) \over 2} \right]=-i \delta ({\bf r}- {\bf r}')
\label{commu1}
\end{eqnarray}
for the canonical quantization condition, where $\hat{\rho}_e$ is the operator corresponding to $\rho_e$.

We construct the following boson field operators,
\begin{eqnarray}
\hat{\psi}_{e}^{\dagger}({\bf r})\!=\! \left(\hat{\rho}_{e}({\bf r}) \right)^{1/2} e^{i {\hat{\chi}({\bf r}) \over 2}}, \ \hat{\psi}_{e}({\bf r})\!=\! e^{-i {{ \hat{\chi}({\bf r}) }\over 2}}\left( \hat{\rho}_{e}({\bf r}) \right)^{1/2}, \ [\hat{\psi}_{e}({\bf r}),\hat{\psi}_{e}^{\dagger}({\bf r}')]\!=\!\delta({\bf r}\!-\!{\bf r}')
\nonumber
\\
\label{boson1}
\end{eqnarray}

 Using the above boson field operators, we construct the number operators for electrons participating in the collective mode described by $\chi$.
 First, we define boson creation operator $\hat{C}_{\chi}^{\dagger}$ and annihilation operator $\hat{C}_{\chi}$
 \begin{eqnarray}
 \hat{C}^{\dagger}_{\chi}=\int_{\cal V} d{\bf r} \hat{\psi}_{e}^{\dagger}({\bf r}) 
, \ \hat{C}_{\chi}=\int_{\cal V} d{\bf r} \hat{\psi}_{e}^{}({\bf r})
\label{Cj}
 \end{eqnarray}
 
 Through the creation and annihilation operators, the phase operator $\hat{X}$ that is conjugate to the number operator $\hat{N}_{\chi}= \hat{C}^{\dagger}_{\chi}  \hat{C}_{\chi}$
 is defined as
  \begin{eqnarray}
 \hat{C}^{\dagger}_{\chi}\!=\!(\hat{N}_{\chi})^{ 1 \over 2} e^{ { i \over 2} \hat{X}}
, \quad \hat{C}_{\chi}\!=\!e^{- { i \over 2} \hat{X}}(\hat{N}_{\chi})^{ 1 \over 2} 
\label{chij}
 \end{eqnarray}
 
We define eigenstates of $\hat{N}_{\chi}$,
    \begin{eqnarray}
\hat{N}_{\chi} | N_{\chi} \rangle =N_{\chi}| N_{\chi} \rangle
 \end{eqnarray}

Then, $e^{\pm { i \over 2} \hat{X}}$ are number changing operators that satisfy
    \begin{eqnarray}
    e^{\pm { i \over 2} \hat{X}} | N_{\chi} \rangle =  | N_{\chi} \pm 1 \rangle
 \end{eqnarray}
 in analogous to Eq.~(\ref{numberOP1}) or Eq.~(\ref{commchi2}).

\section{The Pairing Interaction Using Number Changing Operators $e^{\pm i \hat{X}}$ }
\label{section6}

Using $e^{\pm { i }\hat{X}}$, the interaction part of the Hamiltonian can be written as
\begin{eqnarray}
H_{\rm int}= \sum_{{\bf k} {\bm \ell}} V_{{\bf k} {\bm \ell}}c^{\dagger}_{{\bf k} \uparrow}c^{\dagger}_{-{\bf k} \downarrow} e^{- { i }\hat{X}}   e^{{ i }\hat{X}}c_{-{\bm \ell} \downarrow}c_{{\bm \ell} \uparrow}
\end{eqnarray}
which can be transformed to a mean-field version
\begin{eqnarray}
H^{\rm MF}_{\rm int}&=& \sum_{{\bf k} {\bm \ell}} V_{{\bf k} {\bm \ell}}
\Big[ \langle c^{\dagger}_{{\bf k} \uparrow}c^{\dagger}_{-{\bf k} \downarrow} e^{-  i \hat{X}}    \rangle
e^{i\hat{X}}c_{-{\bm \ell} \downarrow}c_{{\bm \ell} \uparrow}+
c^{\dagger}_{{\bf k} \uparrow}c^{\dagger}_{-{\bf k} \downarrow} e^{- i\hat{X}}   
\langle e^{i \hat{X}}c_{-{\bm \ell} \downarrow}c_{{\bm \ell} \uparrow} \rangle
- \langle
c^{\dagger}_{{\bf k} \uparrow}c^{\dagger}_{-{\bf k} \downarrow} e^{- i\hat{X}} \rangle
\langle
  e^{i \hat{X}}c_{-{\bm \ell} \downarrow}c_{{\bm \ell} \uparrow} \rangle
\Big]
\nonumber
\\
&=& \!-\! \sum_{{\bf k}} \Big[ 
\Delta_{\bf k} c^{\dagger}_{{\bf k} \uparrow}c^{\dagger}_{-{\bf k} \downarrow} e^{- i \hat{X}}   
\!+\! \Delta_{\bf k} e^{ i \hat{X}}c_{-{\bf k} \downarrow}c_{{\bf k} \uparrow}
\!-\! \Delta_{\bf k} \langle
c^{\dagger}_{{\bf k} \uparrow}c^{\dagger}_{-{\bf k} \downarrow} e^{- i \hat{X}} \rangle
\Big]
\label{MF}
\end{eqnarray}
where
\begin{eqnarray}
\Delta_{\bf k}=-\sum_{{\bm \ell}} V_{{\bf k} {\bm \ell}}\langle e^{i \hat{X}}c_{-{\bm \ell} \downarrow}c_{{\bm \ell} \uparrow} \rangle=\Delta_{\rm BCS}
\end{eqnarray}
with employing $V_{{\bf k} {\bm \ell}}< -g$.

We can use the following particle number conserving version of Bogoliubov operators to diagonalize the mean field Hamiltonian \cite{Bogoliubov58,Valatin},
\begin{eqnarray}
\gamma_{{\bf k} 0}&=&u_{k}  e^{ {i \over2}\hat{X}}c_{{\bf k} \uparrow}- v_{k}  c^{\dagger}_{-{\bf k} \downarrow} e^{- {i \over2}\hat{X}}
\nonumber
\\
\gamma_{-{\bf k} 1} &=& u_{k} e^{ {i \over2}\hat{X}}c_{-{\bf k} \downarrow}+ v_{k}  c^{\dagger}_{{\bf k} \uparrow} e^{- {i \over2}\hat{X}}
\label{Bogoliubov1}
\end{eqnarray}
where the operator $u_{k}$ and $v_{ k}$ are given in Eq.~(\ref{uv}). The particle number conserving version of Bogoliubov operators  
$\gamma_{{\bf k} 0}$, $\gamma^{\dagger}_{{\bf k} 0}$, $\gamma_{-{\bf k} 1}$, and $\gamma^{\dagger}_{-{\bf k} 1}$ satisfy the fermion commutation relations.

Reversely, we also have
\begin{eqnarray}
c_{{\bf k} \uparrow}&=&e^{ -{i \over2}\hat{X}} \left( u_{k} \gamma_{{\bf k} 0}  +v_{k} \gamma^{\dagger}_{-{\bf k} 1}  \right) 
\nonumber
\\
c_{-{\bf k} \downarrow}^{\dagger}&=&  \left(-v_{k} \gamma_{{\bf k} 0}  + u_{k}  \gamma^{\dagger}_{-{\bf k} 1} \right)e^{ {i \over2}\hat{X}}
\label{BogoliubovR}
\end{eqnarray}
In order that $c_{{\bf k} \sigma}$ and $c^{\dagger}_{{\bf k} \sigma}$ satisfy the fermion commutation relations, we require that
$\gamma_{{\bf k} 0}$, $\gamma^{\dagger}_{{\bf k} 0}$, $\gamma_{-{\bf k} 1}$, and $\gamma^{\dagger}_{-{\bf k} 1}$ commute with $e^{\pm {i \over2}\hat{X}}$.

Then, the Hamiltonian is cast in the following form
\begin{eqnarray}
H^{\rm MF}= \sum_{\bf k}E_{\bf k}[ \gamma^{\dagger}_{{\bf k} 0} \gamma_{{\bf k} 0} +\gamma^{\dagger}_{{\bf k} 1}\gamma_{{\bf k} 1}]
+ \sum_{\bf k}\left(\xi_0({\bf k}) -  E_{\bf k}+ \Delta_{\rm BCS} \langle
c^{\dagger}_{{\bf k} \uparrow}c^{\dagger}_{-{\bf k} \downarrow} e^{- { i \over 2}\hat{X}} \rangle \right) 
\nonumber
\\
\end{eqnarray}

The ground state for $N$ electron system is the vacuum of $\gamma_{{\bf k} 0}$ and $\gamma_{{\bf k} 1}$,
\begin{eqnarray}
\gamma_{{\bf k} 0} |{\rm Gnd}(N)\rangle =0, \quad \gamma_{{\bf k} 1} |{\rm Gnd}(N)\rangle =0
\end{eqnarray}
 given by
\begin{eqnarray}
|{\rm Gnd}(N)\rangle=\prod_{\bf k} \left( u_{\bf k} + v_{\bf k}  c^{\dagger}_{{\bf k} \uparrow}c^{\dagger}_{-{\bf k} \downarrow} e^{- {i}\hat{X}}\right) |{\rm Cnd} (N)\rangle
\end{eqnarray}
where $|{\rm Cnd } (N)\rangle$ is defined as the state vector for the condensate state given by
\begin{eqnarray}
|{\rm Cnd} (N)\rangle=e^{ i{ N\over2}\hat{X}} |{\rm vac} \rangle
\label{CndBCS}
\end{eqnarray}
analogously to Eq.~(\ref{Cnd0}). 

 In the limit of $\Delta_{\rm BCS} \rightarrow 0$, the ground state becomes the normal state given by
\begin{eqnarray}
{\rm const.}\prod_{k \le k_{F}}  c^{\dagger}_{-{\bf k} \downarrow} c^{\dagger}_{{\bf k} \uparrow}|{\rm vac} \rangle
\end{eqnarray}

\section{ What Is $|{\rm Cnd}(N)\rangle$ ?}
\label{section7}

Let us identify $|{\rm Cnd}(N)\rangle$ in Eq.~(\ref{CndBCS}) in this section. For that purpose we introduce the Berry connection for the electron wave function by following the arguments in Section~\ref{section2}.

We denote the total wave function for the electron system by $\Psi$. We define $|n_{\Psi}({\bf r}) \rangle$ from $\Psi$,
 \begin{eqnarray}
\langle s, {\bf x}_{2}, \cdots, {\bf x}_{M_e} |n_{\Psi}({\bf r},t) \rangle = |C_{\Psi}({\bf r} ,t)|^{-{\frac 1 2}} \Psi({\bf r}s, {\bf x}_{2}, \cdots, {\bf x}_{M_e},t) 
\end{eqnarray}
where ${\bf x}=({\bf r},s)$ ($s$ is the spin coordinate)
with regarding ${\bf r}$ as parameters; $|C_{\Psi}({\bf r} ,t)|$ is the normalization constant given by
\begin{eqnarray}
|C_{\Psi}({\bf r} ,t)|=\int ds d{\bf x}_{2} \cdots d{\bf x}_{M_e}\Psi({\bf r} s, {\bf x}_{2}, \cdots)\Psi^{\ast}({\bf r} s, {\bf x}_{2}, \cdots)
\end{eqnarray}

Then, we obtain the Berry connection and introduce $\chi$ 
\begin{eqnarray}
 {\bf A}_{\Psi}^{\rm MB}=-i \langle n_{\Psi}({\bf r},t)| \nabla_{\bf r} |n_{\Psi}({\bf r},t) \rangle= -{ 1 \over 2} \nabla \chi
\end{eqnarray}
as in Eq.~(\ref{varphi1}).

We can construct a currentless wave function $\Psi_0$ for the current operator associated with $K_0$
\begin{eqnarray}
\Psi_0 ({\bf x}_1, \cdots, {\bf x}_{N},t)=\Psi ({\bf x}_1, \cdots, {\bf x}_{N},t)\exp\left(- i \sum_{j=1}^{N} \int_{0}^{{\bf r}_j} {\bf A}_{\Psi}^{\rm MB}({\bf r}',t) \cdot d{\bf r}' \right)
\end{eqnarray}
and express $\Psi ({\bf x}_1, \cdots, {\bf x}_{N},t)$ as
 \begin{eqnarray}
\Psi ({\bf x}_1, \cdots, {\bf x}_{N},t)=\Psi_0({\bf x}_1, \cdots, {\bf x}_{N},t)\exp\left(- i \sum_{j=1}^{N} { 1 \over 2} \chi ({\bf r}_j, t) \right)
\end{eqnarray}
using the currentless wave function $\Psi_0$ as in Eq.~(\ref{f}).

Then, from the Lagrangian analogous to Eq.~(\ref{L}) but constructed using $\Psi$, we obtain the conjugate momentum for $\chi$,
\begin{eqnarray}
p_{\chi}=\hbar \rho_e/2
\label{momentumchi}
\end{eqnarray}
where $\rho_e$ is the number density of the electrons. 
Consequently, we have the canonical quantization condition in Eq.~(\ref{commu1}).

We associate $|{\rm Cnd}(N)\rangle$ with the following wave function
\begin{eqnarray}
\Psi_0({\bf x}_1, \cdots, {\bf x}_{N},t)=\langle {\bf x}_1, \cdots, {\bf x}_{N} |{\rm Cnd}(N)\rangle
\label{Cnd}
\end{eqnarray}
Actually, we do not need the explicit form for $|{\rm Cnd}(N)\rangle$; the knowledge of ${\bf A}^{\rm MB}_{\Psi}$ is sufficient
for the following calculations.

\section{Josephson Tunneling in the BCS Model}
\label{section8}

Let us consider the observable effect of the number changing operators for the number of particles in the collective mode in this section.
For this purpose we consider the Josephson tunneling \cite{Josephson62}. 

We denote two superconductors in the Josephson junction as $S_L$ and $S_R$. The zeroth order junction state is a product of an $S_L$ state given by
\begin{eqnarray}
|{\rm Gnd}_L(N_L)\rangle=\prod_{{\bf k}_L} \left( u_{k_L} + v_{k_L}  c^{\dagger}_{{\bf k}_L\uparrow}c^{\dagger}_{-{\bf k}_L \downarrow} e^{- {i}\hat{\chi}_L}\right) |{\rm Cnd}_L(N_L)\rangle
\end{eqnarray}
and an $S_R$ state given by
\begin{eqnarray}
|{\rm Gnd}_R(N_R)\rangle=\prod_{{\bf k}_R} \left( u_{k_R} + v_{k_R}  d^{\dagger}_{{\bf k}_R\uparrow}d^{\dagger}_{-{\bf k}_R \downarrow} e^{- {i}\hat{\chi}_R}\right) |{\rm Cnd}_R(N_R)\rangle
\end{eqnarray}
where parameters with subscripts $L$ and $S$ are those for electrons in $S_L$ and $S_R$, respectively; and $c_{{\bf k}_L\sigma}$ and $c^{\dagger}_{{\bf k}_L \sigma}$ denote annihilation and creation operators for electrons in $S_L$, respectively ($d_{{\bf k}_R\sigma}$ and $d^{\dagger}_{{\bf k}_R \sigma}$ denote annihilation and creation operators for electrons in $S_R$, respectively).

The number changing operators $e^{- {i}\hat{\chi}_j}, \ j=L, R$ are defined as follows: First, we construct the number operators for electrons participating in the collective mode described by $\chi$ in the two superconducting regions as
 \begin{eqnarray}
 \hat{C}^{\dagger}_{\chi_j}=\int_{S_j} d{\bf r} \hat{\psi}_{e}^{\dagger}({\bf r}) 
, \ \hat{C}_{\chi_j}=\int_{S_j} d{\bf r} \hat{\psi}_{e}^{}({\bf r}),  \quad \hat{N}_{\chi_j}= \hat{C}^{\dagger}_{\chi_j}  \hat{C}_{\chi_j}
 \end{eqnarray}
 
 Then, the phase operators $\hat{\chi}_j$ that are conjugate to the number operators $\hat{N}_{\chi_j}$
 are defined through the relations
  \begin{eqnarray}
 \hat{C}^{\dagger}_{\chi_j}\!=\!(\hat{N}_{\chi_j})^{ 1 \over 2} e^{ { i \over 2} \hat{\chi}_j}
, \quad \hat{C}_{\chi_j}\!=\!e^{- { i \over 2} \hat{\chi}_j}(\hat{N}_{\chi_j})^{ 1 \over 2} 
\label{chij}
 \end{eqnarray}
 
 The number changing operators are given by $e^{\pm {i}\hat{\chi}_j}$, which satisfy the relations
    \begin{eqnarray}
 e^{\pm {i}\hat{\chi}_j}| N_{\chi_j} \rangle = e^{\pm {i}{\chi}_j} | N_{\chi_j} \pm 1 \rangle
 \label{phaseJJ}
 \end{eqnarray}
Note that the phase factors $e^{\pm {i}{\chi}_j}$ are introduced to take into account the existence of the phase difference between
$|{\rm Cnd}_L(N_L)\rangle$ and $|{\rm Cnd}_R(N_R)\rangle$ arising from Eq.~(\ref{CndBCS}) at the different superconductors.

Now, we consider the standard electron transfer Hamiltonian between $S_L$ and $S_R$ 
\begin{eqnarray}
H_{LR}=-\sum_{{\bf k}_L, {\bf k}_R, \sigma} T_{{\bf k}_L {\bf k}_R} \left(  c^{\dagger}_{{\bf k}_L \sigma} d_{{\bf k}_R \sigma}+d^{\dagger}_{{\bf k}_R\sigma} c_{{\bf k}_L \sigma}  \right)
\label{transfer1}
\end{eqnarray}
where $T_{{\bf k}_L {\bf k}_R} $ is assumed to be real.

It is rewritten using the Boboliubov operators defined analogously to Eq.~(\ref{Bogoliubov1}),
\begin{eqnarray}
H_{LR}&=&-\sum_{{\bf k}_L, {\bf k}_R} T_{{\bf k}_L {\bf k}_R} e^{ { i \over 2}(\hat{\chi}_L-\hat{\chi}_R)}\Big[
(u_{k_L}\gamma^{\dagger}_{{\bf k}_L 0_L} +v_{k_L}\gamma_{-{\bf k}_L 1_L} ) (u_{k_R}\gamma_{{\bf k}_R 0_R} +v_{k_R}\gamma^{\dagger}_{-{\bf k}_R 1_R} ) 
\nonumber
\\
&+&(-v_{k_L}\gamma_{{\bf k}_L 0_L} +u_{k_L}\gamma^{\dagger}_{-{\bf k}_L 1_L} ) (-v_{k_R}\gamma^{\dagger}_{{\bf k}_R 0_R} +u_{k_R}\gamma_{-{\bf k}_R 1_R} ) 
\Big]+\mbox{h.c.}
\end{eqnarray}
by replacing the electron creation and annihilation operators; labels``$L$'' and ``$R$'' refer to quantities for $S_L$ and $S_R$, respectively. 

From the second order perturbation, the effective interaction Hamiltonian that will act on products state $|{\rm Gnd}_L(N_L)\rangle|{\rm Gnd}_R(N_R)\rangle$ is calculated as
\begin{eqnarray}
H_{LR}{1 \over {E_0 - H_0}} H_{LR} &\approx& - \sum_{{\bf k}_R, {\bf k}_L, {\bf k}_R', {\bf k}_L'}T_{{\bf k}_L {\bf k}_R}^2 
\left[ e^{ {i \over 2} (\hat{\chi}_L-\hat{\chi}_R)}v_{k_L}u_{k_R}(\gamma_{-{\bf k}_L 1_L}\gamma_{-{\bf k}_R 0_R}-\gamma_{{\bf k}_L 0_L}\gamma_{-{\bf k}_R 1_R})+ (L \leftrightarrow R) 
\right]
\nonumber
\\
&\times& {1 \over {E_{k_R}+E_{k_L}}}
\left[ e^{ {i \over 2} (\hat{\chi}_L-\hat{\chi}_R)}v_{k'_L}u_{k'_R}(\gamma^{\dagger}_{{\bf k}'_L 0_L}\gamma^{\dagger}_{-{\bf k}'_R 1_R}-\gamma_{-{\bf k}'^{\dagger}_L 1_L}\gamma^{\dagger}_{{\bf k}'_R 0_R})+ (L \leftrightarrow R) 
\right]
\nonumber
\\
&\approx& - \sum_{{\bf k}_R, {\bf k}_L} {{2T_{{\bf k}_L {\bf k}_R}^2 } \over {E_{k_R}+E_{k_L}}}
\left[ v_{k_L}u_{k_L}v_{k_R}u_{k_R}(e^{ {i } (\hat{\chi}_L-\hat{\chi}_R)}+e^{- {i} (\hat{\chi}_L-\hat{\chi}_R)})
+v^2_{k_L}u^2_{k_R} +u^2_{k_L}v^2_{k_R}
\right]
\nonumber
\\
.
\label{Perttransfer3}
\end{eqnarray}

It contains an electron transfer Hamiltonian with the number changing operators, 
\begin{eqnarray}
H_J^{2e}&=&- \sum_{{\bf k}_R, {\bf k}_L} {{2T_{{\bf k}_L {\bf k}_R}^2 } \over {E_{k_R}+E_{k_L}}}
\left[ v_{k_L}u_{k_L}v_{k_R}u_{k_R}(e^{ {i } (\hat{\chi}_L-\hat{\chi}_R)}+e^{- {i} (\hat{\chi}_L-\hat{\chi}_R)})\right]
\nonumber
\\
&=&
- \sum_{{\bf k}_R, {\bf k}_L} {{T_{{\bf k}_L {\bf k}_R}^2 } \over {E_{k_R}+E_{k_L}}} { \Delta_L \over E_{{\bf k}_L}}{ \Delta_R \over E_{{\bf k}_R}}
\cos \left( \hat{\chi}_L-\hat{\chi}_R \right)
\label{transfer3}
\end{eqnarray}
 where $\Delta_L$ and $\Delta_R$ are energy gaps in $S_L$ and $S_R$, respectively.
 The above transfer Hamiltonian indicates that tunneling of the condensate occurs through the difference of the phase operators  $\left( \hat{\chi}_L-\hat{\chi}_R \right)$.
It transfers two electrons, giving the interpretation that the tunneling current is due to the electron-pair flow; however, from the view point of the present theory, Josephson tunneling current should be considered as due to the collective mode $\nabla \chi$.
In this respect, it is worth noting that a doubt has been raised in the attribution of the Josephson tunneling to the electron-pair flow \cite{Koizumi2011,HKoizumi2015}. We will come back to this problem in Section \ref{section11}.

\section{$\Psi_s$ in the BCS Model}
\label{section9}

In order to describe the supercurrent, spatial variation of $\chi$ needs to be included.  
We have obtained the result for two sites, $S_L$ and $S_R$, in the previous section; we will extend it to the many-site case in this section.

We express the ground state for the many-site case as 
\begin{eqnarray}
|{\rm Gnd}({\bf r}; N)\rangle=\prod_{\bf k} \left( u_{\bf k}({\bf r}) + v_{\bf k}({\bf r})  c^{\dagger}_{{\bf k} \uparrow}c^{\dagger}_{-{\bf k} \downarrow} e^{- {i}\hat{\chi}({\bf r}) }\right) |{\rm Cnd} (N)\rangle
\label{newGnd}
\end{eqnarray}
where ${\bf r}$ denotes the centers of the coarse-grained cells.
 In the following, we treat ${\bf r}$ as a continuous variable using spatially varying parameters $u_k({\bf r}) $, $v_k({\bf r})$, and $e^{- {i}\hat{\chi}({\bf r}) }$.

The requirement that the ground state in Eq.~(\ref{BCSr}) is obtained from Eq.~(\ref{newGnd}) in a certain approximation
indicates that the phase of the condensate state vector at ${\bf r}$ satisfies
\begin{eqnarray}
e^{- {i}\hat{\chi}({\bf r}) }|{\rm Cnd} (N)\rangle=e^{- {i}{\chi}({\bf r}) }|{\rm Cnd} (N-2)\rangle
\label{chiphase}
\end{eqnarray}
as in Eq.~(\ref{phaseJJ}).

Then, Eq.~(\ref{BCSr}) can be viewed as an approximation to Eq.~(\ref{newGnd}) in which $e^{- {i}\hat{\chi}({\bf r}) }$ is replaced by $e^{- {i}{\chi}({\bf r}) }$; and $|{\rm Cnd} (N)\rangle$ is also replaced by $|{\rm vac} \rangle$ to take into account the reduction of the number of particles by $e^{- {i}\hat{\chi}({\bf r}) }$, which is absent for $e^{- {i}{\chi}({\bf r}) }$.

The electron field operators $\hat{\Psi}_{\sigma}({\bf r})=\sum_{\bf k} e ^{ i {\bf k} \cdot {\bf r}} / \sqrt{\cal V} c_{{\bf k} \sigma}, \ \sigma=\uparrow, \downarrow$, are now modified to
\begin{eqnarray}
\hat{\Psi}^{\rm BCS}_{\uparrow}({\bf r}) &=&{ 1 \over \sqrt{\cal V}} \sum_{\bf k} e^{-i{ 1 \over 2} \hat{\chi}({\bf r})} e^{i{\bf k} \cdot {\bf r}} \left( u_k({\bf r})\gamma_{{\bf k} 0} +   v_k({\bf r}) \gamma^{\dagger}_{-{\bf k} 1}
\right)
\\
\hat{\Psi}^{\rm BCS}_{\downarrow}({\bf r}) &=&{ 1 \over \sqrt{\cal V}}\sum_{\bf k} e^{-i{ 1 \over 2}\hat{\chi}({\bf r})} 
e^{-i{\bf k} \cdot {\bf r}}\left( -v_k({\bf r})\gamma^{\dagger}_{{\bf k} 0} +   u_k({\bf r}) \gamma_{-{\bf k} 1} \right)
\end{eqnarray}
considering Eq.~(\ref{BogoliubovR}) and incorporating the spatial variations of $u_k$, $v_k$, and $e^{- {i}\hat{\chi} }$.

Then, $\Psi_s$ for the present model, $\Psi^{\rm BCS}_s$, is given by
\begin{eqnarray}
\Psi^{\rm BCS}_s&=& \langle {\rm Gnd}({\bf r}; N)| e^{i \hat{X} }\hat{\Psi}^{\rm BCS}_{\downarrow}({\bf r}) \hat{\Psi}^{\rm BCS}_{\uparrow}({\bf r}) |{\rm Gnd}({\bf r}; N)\rangle
=
e^{-i {\chi}({\bf r})} { 1 \over {\cal V}}\sum_{\bf k} u_k({\bf r}) v_k({\bf r}) 
\nonumber
\\
&=&
 {{ \Delta_{\rm BCS}({\bf r})} \over {g {\cal V}}}   e^{-i{\chi}({\bf r})}
\end{eqnarray}
in analogous to Eq.~(\ref{Psis}).

\section{Bogoliubov-de Gennes Equations}
\label{section10}

For inhomogeneous superconductors, a powerful formalism is the Bogoliubov-de Gennes formalism \cite{deGennes}. In this section, we will extend the present formalism to the inhomogeneous case by following the Bogoliubov-de Gennes formalism.

The effective Hamiltonian of the Bogoliubov-de Gennes formalism is now given by
\begin{eqnarray}
{\cal H}_{\rm eff}=\int d{\bf r} \Big[
\sum_{\sigma} \left( \hat{\Psi}^{\dagger}_{\sigma}({\bf r}) {\cal H}_e \hat{\Psi}_{\sigma}({\bf r})
+U({\bf r})  \hat{\Psi}^{\dagger}_{\sigma}({\bf r}) \hat{\Psi}_{\sigma}({\bf r}) \right) +\Delta({\bf r})e^{-i\hat{\chi}({\bf r})} \hat{\Psi}^{\dagger}_{\uparrow}({\bf r}) \hat{\Psi}^{\dagger}_{\downarrow}({\bf r}) + \Delta^{\ast}({\bf r})e^{i\hat{\chi}({\bf r})} \hat{\Psi}_{\downarrow}({\bf r}) \hat{\Psi}_{\uparrow}({\bf r}) 
\Big]
\nonumber
\\
\label{Heff}
\end{eqnarray}
where
\begin{eqnarray}
\hat{\Psi}_{\uparrow}({\bf r}) &=&\sum_n e^{-i{ 1 \over 2}\hat{\chi}({\bf r})} ( \gamma_{n \uparrow}u_n({\bf r}) - \gamma^{\dagger}_{n \downarrow}v^{\ast}_n({\bf r}) 
)
\\
\hat{\Psi}_{\downarrow}({\bf r}) &=&\sum_n e^{-i{ 1 \over 2}\hat{\chi}({\bf r})} ( \gamma_{n \downarrow}u_n({\bf r}) + \gamma^{\dagger}_{n \uparrow}v^{\ast}_n({\bf r}) 
)
\end{eqnarray}
and
\begin{eqnarray}
{\cal H}_e &=& { 1 \over {2m}} 
\left( -i\hbar \nabla -q{\bf A}^{\rm em} \right)^2 +U_0({\bf r})-{\cal E}_F
\\
U({\bf r})&=&-g\langle   \hat{\Psi}^{\dagger}_{\uparrow }({\bf r}) \hat{\Psi}_{\uparrow}({\bf r}) \rangle 
=- g\langle   \hat{\Psi}^{\dagger}_{\downarrow }({\bf r}) \hat{\Psi}_{\downarrow}({\bf r}) \rangle 
\\
\Delta({\bf r})&=&-g \langle   e^{i \hat{\chi}({\bf r})}\hat{\Psi}_{\downarrow }({\bf r}) \hat{\Psi}_{\uparrow}({\bf r}) \rangle 
=g \langle e^{i \hat{\chi}({\bf r})} \hat{\Psi}_{\uparrow }({\bf r}) \hat{\Psi}_{\downarrow}({\bf r}) \rangle 
\end{eqnarray}

In the single-particle Hamiltonian ${\cal H}_e$, a potential $U_0({\bf r})$ and the electromagnetic vector potential ${\bf A}^{\rm em}$ are included; $q$ is the electron charge $q=-e$.

The Bogoliubov operators $\gamma_{n \sigma}$ and $\gamma^{\dagger}_{n \sigma}$ are those conserves particle numbers like those in Eq.~(\ref{Bogoliubov1}). 
They obey fermion commutation relations, and are chosen to satisfy
\begin{eqnarray}
\left[ {\cal H}_{\rm eff}, \gamma_{n \sigma } \right] &=&-\epsilon_n \gamma_{n \sigma}
\\
\left[{\cal H}_{\rm eff}, \gamma^{\dagger}_{n \sigma } \right] &=&\epsilon_n \gamma^{\dagger}_{n \sigma}
\end{eqnarray}

Using Eq.~(\ref{Heff}) and commutation relations for $\hat{\Psi}^{\dagger}_{\sigma }({\bf r})$ and $\hat{\Psi}_{\sigma }({\bf r})$,
the following relations are obtained
\begin{eqnarray}
\left[\hat{\Psi}_{\uparrow }({\bf r}) , {\cal H}_{\rm eff} \right] &=&
\left[{\cal H}_e + U({\bf r}) \right] \hat{\Psi}_{\uparrow }({\bf r})+\Delta({\bf r}) e^{-i \hat{\chi}({\bf r})}\hat{\Psi}^{\dagger}_{\downarrow }({\bf r})
\label{deG1}
\\
\left[\hat{\Psi}_{\downarrow }({\bf r}) , {\cal H}_{\rm eff} \right] &=&
\left[{\cal H}_e + U({\bf r}) \right] \hat{\Psi}_{\downarrow }({\bf r})-\Delta({\bf r}) e^{-i \hat{\chi}({\bf r})}\hat{\Psi}^{\dagger}_{\uparrow }({\bf r})
\label{deG2}
\end{eqnarray}

By taking into account the relation in Eq.~(\ref{chiphase}), we can replace $e^{-i{ 1 \over 2}\hat{\chi}({\bf r})}$ with $e^{-i{ 1 \over 2}{\chi}({\bf r})}$. Then, we obtain the following system of equations,
\begin{eqnarray}
\epsilon_n u_n({\bf r})&=&
\left[\bar{\cal H}_e + U({\bf r}) \right] u_n({\bf r})+\Delta ({\bf r})v_n({\bf r})
\\
\epsilon_n v_n({\bf r})&=&-
\left[\bar{\cal H}^{\ast}_e + U({\bf r}) \right] v_n({\bf r})+\Delta^{\ast}({\bf r})u_n({\bf r})
\end{eqnarray}
where
\begin{eqnarray}
U({\bf r})&=&-g\sum_n |v_n({\bf r})|^2
\\
\Delta({\bf r})&=&g\sum_n v^{\ast}_n({\bf r})  u_n({\bf r})
\\
\bar{\cal H}_e &=& { 1 \over {2m}} 
\left( -i\hbar \nabla -q{\bf A}^{\rm eff} \right)^2 +U_0({\bf r})-E_F
\end{eqnarray}
with
\begin{eqnarray}
{\bf A}^{\rm eff} ={\bf A}^{\rm em} +{\hbar \over {2q}} \nabla \chi
\label{Aeff}
\end{eqnarray}

In this system equations ${\bf A}^{\rm eff}$ appears instead of ${\bf A}^{\rm em}$. The effective vector potential ${\bf A}^{\rm eff}$ is gauge invariant since the gauge degree of freedom in ${\bf A}^{\rm em}$ is compensated by that in $\nabla \chi$ which arises from the wave function. 

Now $\Psi_s$ is given by
\begin{eqnarray}
\Psi_s&=& \langle  e^{i \hat{X} }\hat{\Psi}_{\uparrow}({\bf r}) \hat{\Psi}_{\downarrow}({\bf r}) \rangle
=
e^{-i {\chi}({\bf r})}\sum_{n} u_n({\bf r}) v^{\ast}_n({\bf r}) =g^{-1}\Delta({\bf r}) e^{-i {\chi}({\bf r})}
\label{PsiBdG}
\end{eqnarray}
This is essentially the one derived by Gor'kov \cite{Gorkov}.

\section{Concluding Remarks}
\label{section11}

We have obtained the macroscopic wave function for superconductivity in Eq.~(\ref{PsiBdG}}). The appearance of it corresponds to the stabilization of the collective mode described by $\chi$. It is achieved by the electron-pair formation.
Thus, the superconducting phase transition temperature corresponds to the electron-pair formation temperature.

In the present theory, superconductivity requires $\chi$ in addition to the electron-pairs.
In other words, the supercurrent generation requires an additional ingredient to the BCS theory in accordance with the warning by Bloch \cite{Bloch1966}. In the standard theory, this additional ingredient is supplied by the gauge symmetry breaking brought about by the use of the particle number non-fixed formalism. In contrast to it, it is supplied as the nontrivial Berry connection in the present formalism.

It is now widely believed that supercurrent generation is due to the electron-pair flow due to the observation of the flux quantum $h/2e$ and also due to the observation of the ac Josephson effect \cite{Josephson62}. 
In the present formalism, however, the flux quantum $h/2e$ is rather attributed to the Berry connection in Eq.~(\ref{Aeff}).
Although the ac Josephson effect is believed to confirm that the supercurrent is due to the electron pair flow,
it is notable that a serious misfit was found in the boundary condition employed for the standard derivation and that in the real experimental situation \cite{Koizumi2011,HKoizumi2015}. We consider this problem below.

 In the standard theory, the electron transfer Hamiltonian in Eq.~(\ref{transfer1}) is used, which yields electron-pair transfer Hamiltonian $H^{2e}_J$ in Eq.~(\ref{transfer3}).
Then,  Eq.~(\ref{phi}) is obtained from $\phi$ given by
\begin{eqnarray}
\phi= { {2e} \over \hbar } \int_L^R {\bf A}^{\rm eff} \cdot d {\bf r}
\label{phi2}
\end{eqnarray}
where the gauge invariant vector potential ${\bf A}^{\rm eff}$ in Eq.~(\ref{Aeff}) is used to include ${\bf A}^{\rm em}$, and the integration is performed along a line connecting $S_L$ and $S_R$.

However, in the real experimental situation, the electrons do not just hop between the two superconductors, but they enter from the lead and flow out to the lead. By taking into account the latter effects, it is shown that $\phi$ satisfies Eq.~(\ref{phi}) is not the one in Eq.~(\ref{phi2}), but the following \cite{Koizumi2011,HKoizumi2015},
\begin{eqnarray}
\phi= { e \over \hbar } \int_L^R {\bf A}^{\rm eff} \cdot d {\bf r}
\label{phi1}
\end{eqnarray}

This $\phi$ is obtained if we use the following electron transfer Hamiltonian,
\begin{eqnarray}
H_J^{e}=-T \left(   \hat{C}^{\dagger}_{\chi_L} \hat{C}_{\chi_R} + \hat{C}^{\dagger}_{\chi_R} \hat{C}_{\chi_L}  \right)
\approx 
-2T (N_{\chi_L} N_{\chi_R} )^{1/2} \cos {{ \hat{\chi}_{L} - \hat{\chi}_{R} } \over 2}
\label{transfer2}
\end{eqnarray}
where the operators $(\hat{N}_{\chi_j})^{1/2},  \  j=L,R$  are replaced by their expectation values in the rightmost expression.

Including ${\bf A}^{\rm em}$ and replacing $\hat{\chi}_{j}$, $j=L, R$ by their expectation values, $H_J^{e}$ becomes one with the gauge invariant ${\bf A}^{\rm eff}$, 
\begin{eqnarray}
H_J^{e}
\approx 
-2T (N_{\chi_L} N_{\chi_R} )^{1/2} \cos \left( { e \over \hbar } \int_L^R {\bf A}^{\rm eff} \cdot d {\bf r} \right) 
\end{eqnarray}
which yields $\phi$ in Eq.~(\ref{phi1}). Thus, the electron transfer Hamiltonian relevant for the ac Josephson effect is $H_J^{e}$, rather than $H_J^{2e}$.

Lastly, we would like to mention that the supercurrent is attributed to the collective motion of $\nabla \chi$, rather than electron-pair flow in the present formalism. This suggests that superconductivity will occur without pairing electrons if a nontrivial $\nabla \chi$ exists and stabilized by some means.

\section*{Acknowledgement}
 Part of the present work was conducted during the author's sabbatical stay at Lorentz Institute for theoretical physics, Leiden University, the Netherlands. 
He thanks the members of the institute for their hospitality.


\begin{thebibliography}{10}
\providecommand{\url}[1]{{#1}}
\providecommand{\urlprefix}{URL }
\expandafter\ifx\csname urlstyle\endcsname\relax
  \providecommand{\doi}[1]{DOI \discretionary{}{}{}#1}\else
  \providecommand{\doi}{DOI \discretionary{}{}{}\begingroup
  \urlstyle{rm}\Url}\fi

\bibitem{Anderson66}
P.W. Anderson, Rev. Mod. Phys. \textbf{38}, 298 (1966)

\bibitem{WWW1970}
G.C. Wick, A.S. Wightman, E.P. Wigner, Phys. Rev. D \textbf{1}, 3267 (1970)

\bibitem{Peierls1991}
R.~Peierls, J. Phys. A \textbf{24}, 5273 (1991)

\bibitem{Peierls92}
R.~Peierls, Contemporary Phys. \textbf{33}, 221 (1992)

\bibitem{LeggettBook}
A.J. Leggett, \emph{Quantum Liquids: Bose Condensation And Cooper Pairing in
  Condensed-matter Systems} (Oxford Univ. Press, Oxford, 2006)

\bibitem{Leggett2001}
A.J. Leggett, Rev. Mod. Phys. \textbf{73}, 307 (2001)

\bibitem{Gross}
E.~Gross, Nuovo Cim \textbf{20}, 454 (1961)

\bibitem{Pitaevskii}
L.P. Pitaevskii, Sov. Phys. JETP. 13 (2): \textbf{13}, 451 (1961)

\bibitem{Landau1941}
L.D. Landau, Journal of Physics (USSR) \textbf{5}, 71 (1941)

\bibitem{Bogolubov47}
N.~Bogolubov, J. Phys. U.S.S.R. \textbf{11}, 23 (1947)

\bibitem{GL}
V.L. Ginzburg, L.D. Landau, Zh. Exsp. Teor. Fiz. \textbf{20}, 1064 (1950)

\bibitem{Gorkov}
L.P. Gor'kov, Sov. Phys. JETP \textbf{9}, 1364 (1959)

\bibitem{Nori2017}
X.~Gu, A.F. Kochum, A.~Miranowicz, Y.~Liu, F.~Nori, Phys. Report
  \textbf{718-719}, 1 (2017)

\bibitem{Phase-Angle}
P.~Carruthers, M.M. Nieto, Rev. Mod. Phys. \textbf{40}, 411 (1968)

\bibitem{Fujikawa2004}
K.~Fujikawa, H.~Suzuki, \emph{Path Integrals and Quantum Anomalies} (Oxford
  Univ. Press, 2004)

\bibitem{Berry}
M.V. Berry, Proc. Roy. Soc. London Ser. A \textbf{391}, 45 (1984)

\bibitem{BMKNZ}
A.~Bohm, A.~Mostafazadeh, H.~Koizumi, Q.~Niu, J.~Zwanziger, \emph{The Geometric
  Phase in Quantum Systems} (Springer, Heidelberg, 2003)

\bibitem{Koonin1976}
A.K. Kerman, S.E. Koonin, Ann. Phys. \textbf{100}, 332 (1976)

\bibitem{Anderson1952}
P.W. Anderson, Phys. Rev \textbf{86}, 694 (1952)

\bibitem{Kubo1952}
R.~Kubo, Phys. Rev. \textbf{87}, 568 (1952)

\bibitem{Anderson1958b}
P.W. Anderson, Phys. Rev. \textbf{112}, 1900 (1958)

\bibitem{BCS1957}
J.~Bardeen, L.N. Cooper, J.R. Schrieffer, Phys. Rev. \textbf{108}, 1175 (1957)

\bibitem{Bogoliubov58}
N.N. Bogoliubov, Soviet Physics JETP \textbf{34}, 41 (1958)

\bibitem{Valatin}
J.G. Valatin, Nuovo Cimento \textbf{7}, 843 (1958)

\bibitem{Josephson62}
B.D. Josephson, Phys. Lett. \textbf{1}, 251 (1962)

\bibitem{Koizumi2011}
H.~Koizumi, J. Supercond. Nov. Magn. \textbf{24}, 1997 (2011)

\bibitem{HKoizumi2015}
H.~Koizumi, M.~Tachiki, J. Supercond. Nov. Magn. \textbf{28}, 61 (2015)

\bibitem{deGennes}
P.G. de~Gennes, \emph{Superconductivity of Metals and Alloys} (W. A. Benjamin,
  Inc., 1966)

\bibitem{Bloch1966}
F.~Bloch, Physics Today \textbf{19}(5), 27 (1966)

\end{thebibliography}
\end{document}